\documentclass[11pt]{article}

\usepackage[utf8]{inputenc}
\usepackage[T1]{fontenc}
\usepackage{graphicx}
\usepackage[margin=1in]{geometry}
\usepackage{ragged2e}
\usepackage{gensymb}
\usepackage[version=3]{mhchem} 
\usepackage{siunitx}
\DeclareSIUnit{\litre}{l}
\usepackage{hyperref}
\usepackage{physics}
\usepackage{braket}
\usepackage{float} 
\usepackage{authblk} 

\title{A Compact Incubation Platform for Long-Term Cultivation of Biological
Samples for Nitrogen-Vacancy Center Widefield Microscopy}

\author[1,*]{Andre Pointner}
\author[2]{Daniela Thalheim}
\author[2]{Sarah Belasi}
\author[3]{Lukas Heinen}
\author[3]{Lucas R. Carnell}
\author[3]{Christina Janko}
\author[3]{Rainer Tietze}
\author[3]{Christoph Alexiou}
\author[2]{Regine Schneider-Stock}
\author[1,*]{Roland Nagy}

\affil[1]{Institute of Applied Quantum Technologies, Friedrich-Alexander-Universit\"at Erlangen-N\"urnberg, Erlangen, 91052, Germany}
\affil[2]{Experimental Tumor Pathology, Universit\"atsklinikum Erlangen, Friedrich-Alexander-Universit\"at Erlangen-N\"urnberg, Erlangen, 91052, Germany}
\affil[3]{Department of Otorhinolaryngology-Head and Neck Surgery, Section of Experimental Oncology and Nanomedicine (SEON), Else Kroener-Fresenius-Stiftung-Professorship, Universit\"atsklinikum Erlangen, Erlangen, 91052, Germany}
\affil[*]{Email: andre.pointner@fau.de, roland.nagy@fau.de}

\date{}

\begin{document}

\maketitle

\begin{center}
\end{center}

\begin{abstract}
Nitrogen-vacancy (NV) centers in diamond provide a versatile quantum sensing
platform for biological imaging through magnetic field detection, offering
unlimited photostability and the ability to perform long-term observations
without photobleaching or phototoxicity. However, conventional stage-top
incubators are incompatible with the unique requirements for NV widefield
magnetometry to study cellular dynamics. Here, we present a purpose-built
compact incubation platform that maintains precise environmental control of
temperature, CO$_2$ atmosphere, and humidity while accommodating the complex
constraints of NV widefield microscopy. The system employs a 3D-printed
biocompatible chamber with integrated heating elements, temperature control, and
humidified gas flow to create a stable physiological environment directly on the
diamond sensing surface. We demonstrate sustained viability and proliferation of
HT29 colorectal cancer cells over 90 hours of continuous incubation, with
successful magnetic field imaging of immunomagnetically labeled cells after
extended cultivation periods. This incubation platform enables long-term
cultivation and real-time monitoring of biological samples on NV widefield
magnetometry platforms, opening new possibilities for studying dynamic cellular
processes using quantum sensing technologies.
\end{abstract}

\section{Introduction}
\label{sec:introduction}
Cell imaging is an
indispensable technique in modern medical and biological
research, enabling
researchers to decipher complex cellular processes and
quantify detailed
kinetic metrics
\cite{Mullassey2008-xx,Gahlmann2014-sr,Schmidt2021-mf,Lu2023-dm}. While
conventional fluorescence microscopy remains the gold standard for cellular
imaging, it faces inherent limitations including photobleaching of fluorophores
and phototoxicity during extended observation periods. These constraints
restrict the duration and quality of time-lapse studies of dynamic cellular
processes, particularly for applications requiring continuous long-term
monitoring.

NV centers in diamond have emerged as a versatile sensing platform that
addresses many of these limitations. NV centers offer unlimited photostability
and can serve as quantum sensors for multiple physical quantities, including
magnetic fields, temperature, and electric fields
\cite{Zhang2021Toward,Hollendonner2023-yi,Liu2022-xx}. Recent studies have
demonstrated the use of NV-doped nanodiamonds for intracellular sensing of
temperature and reactive oxygen species in living cells, providing novel
insights into cellular processes at the submicron scale
\cite{Schirhagl2014-xx,Kucsko2013-th,PeronaMartinez2020-nd,Fan2025-iv}. For
biological imaging, magnetic field sensing using NV centers combined with
superparamagnetic iron oxide nanoparticles (SPIONs) as magnetic labels provides
a compelling alternative to fluorescence imaging
\cite{Pointner2025-acs,Le_Sage2013-at,Chen2022-cs,Glenn2015-gy,Friedrich2022-sf}.
Magnetic fields penetrate biological matter without significant absorption or
perturbation, enabling deeper tissue imaging and longer observation periods
without photobleaching or phototoxicity. NV widefield microscopy achieves
submicron spatial resolution by combining optical resolution with
immunomagnetic
labeling technology across the entire field-of-view
\cite{Sengottuvel2022-xx}.

Despite these advantages, the unique experimental geometry of NV widefield
magnetometry poses challenges for maintaining viable cell cultures during
measurements. NV widefield microscopy typically employs total internal
reflection (TIR) excitation to minimize the interaction of high-intensity
excitation light with biological samples while maximizing NV center fluorescence
\cite{Le_Sage2013-at}. This inverted illumination geometry is incompatible with
commercial stage-top and whole-microscope incubators designed for conventional
inverted microscopes with objective-based imaging
\cite{Wijewardhane2024-lc,Worcester2025-st}. The resulting lack of environmental
control during measurements limits the ability to perform long-term studies of
cellular dynamics using NV widefield magnetic imaging.

Here, we present a purpose-built compact incubation platform specifically
designed for long-term cultivation of biological samples on a diamond in a NV
widefield microscope. Our system accommodates the TIR illumination geometry
required for NV widefield magnetometry while maintaining precise control of
temperature, CO$_2$ atmosphere, and humidity. We demonstrate sustained cell
viability and proliferation over 90 hours of continuous incubation, enabling the
cultivation of cells under physiologically relevant conditions directly on the
diamond surface, without compromising the magnetic field sensing capabilities of
the NV widefield magnetometry system.

\section{Results and Discussion}
\label{sec:results_and_discussion}

\begin{figure}
\centering
\includegraphics[width=\linewidth]{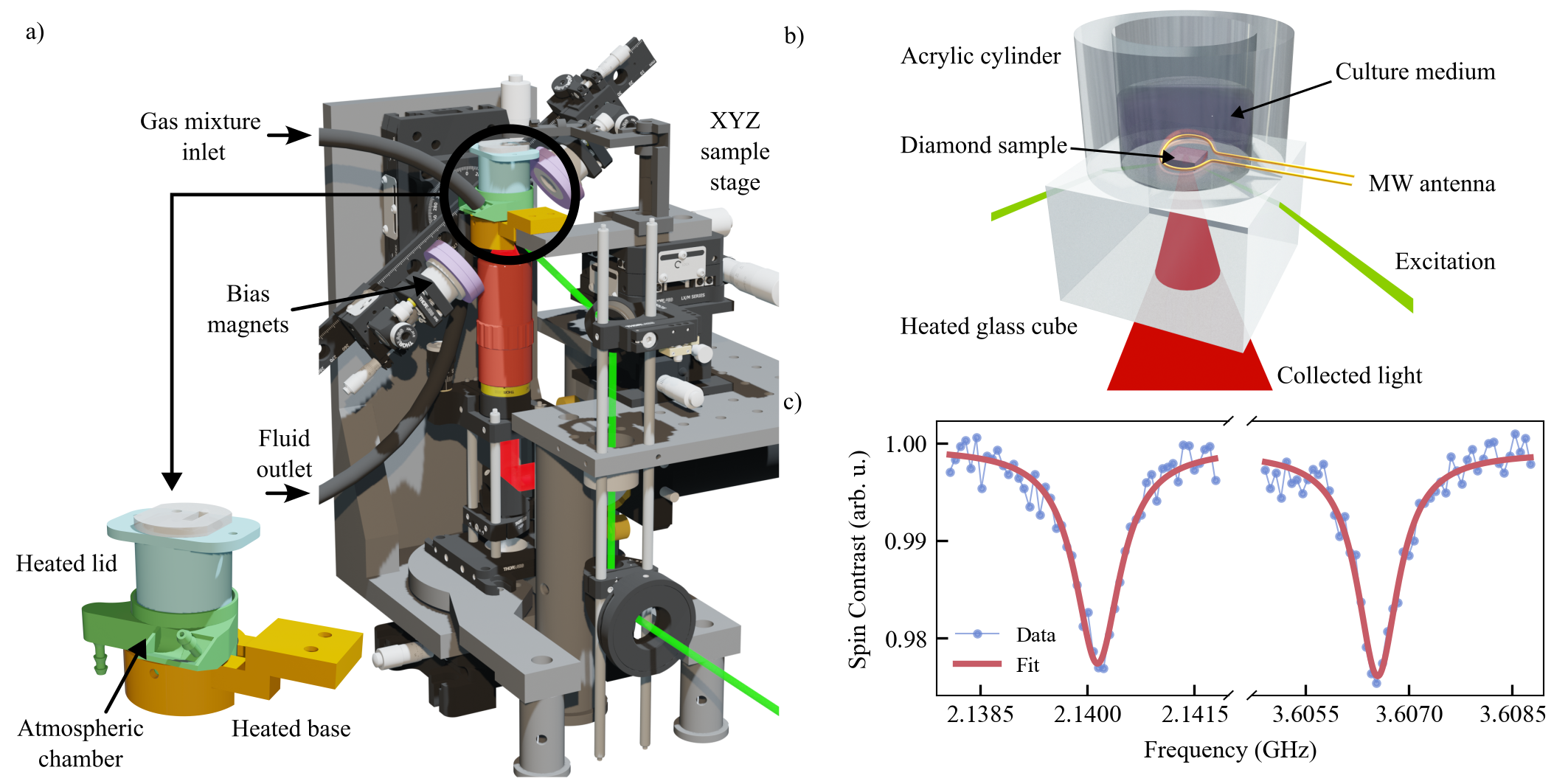}
\caption{a) 3D render of the experimental apparatus. The long working-distance
objective (red) collects light from the diamond sample positioned by the base
heater (orange). The incubation chamber (green) and incubator lid (blue) provide
an isolated atmosphere around the sensing volume. A pair of permanent magnets
(purple) provides a bias magnetic field to prime the electron Zeeman effect and
magnetize the SPIONs in the sample. b) Cross-section render of the TIR
illumination geometry. The excitation laser (green) is coupled into the glass
cube (light gray) and directed into the diamond (red). The TIR at the
diamond-sample interface prevents excitation light from reaching the biological
sample (pink). The microwave antenna is looped around the sample and tilted to
provide strong and homogeneous coupling to the interrogated NV axis. The diamond
and microwave antenna are submerged in the culture medium along with the
cultivated sample. c) ODMR spectrum of a single pixel in the FoV of the camera
demonstrating the dual tone ODMR signal. The microwave signal and the N$^{15}$
hyperfine splitting cause three ODMR resonances from which the center one
observes the most contrast. This feature is evaluated to extract the locally
resolved electron Zeeman effect and subsequently the magnetic field. The
asymmetry of the resonance is attributed to the frequency response of the
microwave antenna.}
\label{fig:1}
\end{figure}

The NV center based widefield magnetic field microscope (see
\autoref{fig:1}a) is built around a [100] bulk diamond sample with a
\SI{1}{\mu m} thick, \SI{1}{ppm} NV layer fabricated by Quantum Diamonds
\cite{quantum-diamonds}. The diamond is mounted on a polished glass cube to
enable TIR excitation of the NV centers, to minimize interaction of the
excitation light with biological samples (see \autoref{fig:1}b)
\cite{Le_Sage2013-at}. An enameled copper wire is positioned around the diamond
sample at an angle approximately \SI{36}{\degree} to provide a uniform
oscillating magnetic field over the field of view (FoV) perpendicular to the
selected NV-axis for spin driving. Two samarium cobalt ring magnets separated by
\SI{70}{\milli\meter} are positioned outside the incubation chamber to provide a
homogeneous static magnetic field of up to \SI{30}{\milli\tesla} along the
selected NV-axis \cite{Bucher2019-nmr}. This magnetic field magnetizes the
SPIONs present in biological samples \cite{Pointner2025-acs} along the sensing
axis. Excitation of the NV centers in the top surface layer of the diamond
crystal is achieved by coupling a \SI{532}{\nano\meter} laser (Verdi G Series,
Coherent) into the side surface of the polished glass cube (as illustrated in
\autoref{fig:1}b). Collimated excitation light is focused and directed through
the glass cube to couple into the bottom diamond surface between
\SIrange{36}{39}{\degree} depending on the chosen FoV. The TIR at the top
diamond surfaces reflects the beam symmetrically away from the biological sample
and out of the glass cube. A long working distance objective (MY20X-824,
Mitutoyo) is used to collect light from a red LED for brightfield imaging and
the NV fluorescence for magnetic field imaging. For brightfield images the full
\SI{652}{\micro\meter} $\times$ \SI{652}{\micro\meter} FoV is used, while the NV
emission is evaluated over a \SI{318.5}{\micro\meter} $\times$
\SI{318.5}{\micro\meter} FoV to achieve a higher saturation of the NV centers
due to a larger laser intensity (\SI{1}{\kilo\watt\per\cm\squared}). The
collected light is filtered by a long-pass filter (FELH0600, Thorlabs, Inc.) to
eliminate scattered excitation light before impinging on a EMCCD camera
(Evolve13, Teledyne) where the signal is 2$\times$2 binned to improve the
signal-to-noise ratio (SNR). A dual-tone microwave signal is used to drive the
NV centers N$^{15}$ hyperfine split ground state spin transitions. The signal is
generated by In-phase and quadrature (IQ) mixing (MMIQ-0218, Marki Microwave) a
constant \SI{1.5}{\MHz} intermediate frequency (IF) signal from an AWG (OPX+,
Quantum Machines) with a local oscillator (LO) signal (SynthNV Pro, Windfreak)
that is swept across the NV resonance frequencies synchronized with the camera
captures, orchestrated by the AWG. The IQ mixing parameters are specifically
tuned to suppress LO leakage and reduce sidebands to a minimum (\SI{-30}{dBm}
suppression threshold) while maintaining the LO+IF and LO-IF images. The
resulting fluorescence images are interleaved with and without microwave driving
enabled to compute the optically detected magnetic resonance (ODMR) contrast
(see \autoref{fig:1}c). The dual tone scheme allows driving both N$^{15}$
hyperfine transitions simultaneously and results in three distinct ODMR
resonances from which the center resonance is evaluated to extract the local
magnetic field based on the electron Zeeman effect. Evaluating every pixel in
the FoV results in a 2D map of the magnetic field sensed by the NV centers along
the interrogated axis. Subsequent background subtraction by a rolling ball
subtraction removes low-frequency spatial noise from the magnetic field images,
revealing the magnetic field patterns generated by the SPIONs in the biological
samples \cite{Pointner2025-acs,Le_Sage2013-at,Glenn2015-gy,Chen2022-cs}.

In order to correlate magnetic field images captured by ODMR, a red LED is
positioned above the incubation chamber to enable brightfield imaging.
Fluorescence imaging capabilities are optionally enabled by integrating a
\SI{470}{\nano\meter} LED illumination source (M470L5, Thorlabs Inc.) into the
emission path through a dichroic mirror (DMLP490L, Thorlabs, Inc.). This allows
the use of common fluorescent dyes to perform vitality assays of the cells
inside the incubator \cite{Le_Sage2013-at,Robertson2019-bl}. We isolate the
fluorescence signals using a bandpass emission filter (MF525-39, Thorlabs,
Inc.)
for green and the long-pass filter used for the NV-emission for red
channels
during fluorescence imaging.

\begin{figure}[H]
\centering
\includegraphics[width=\linewidth]{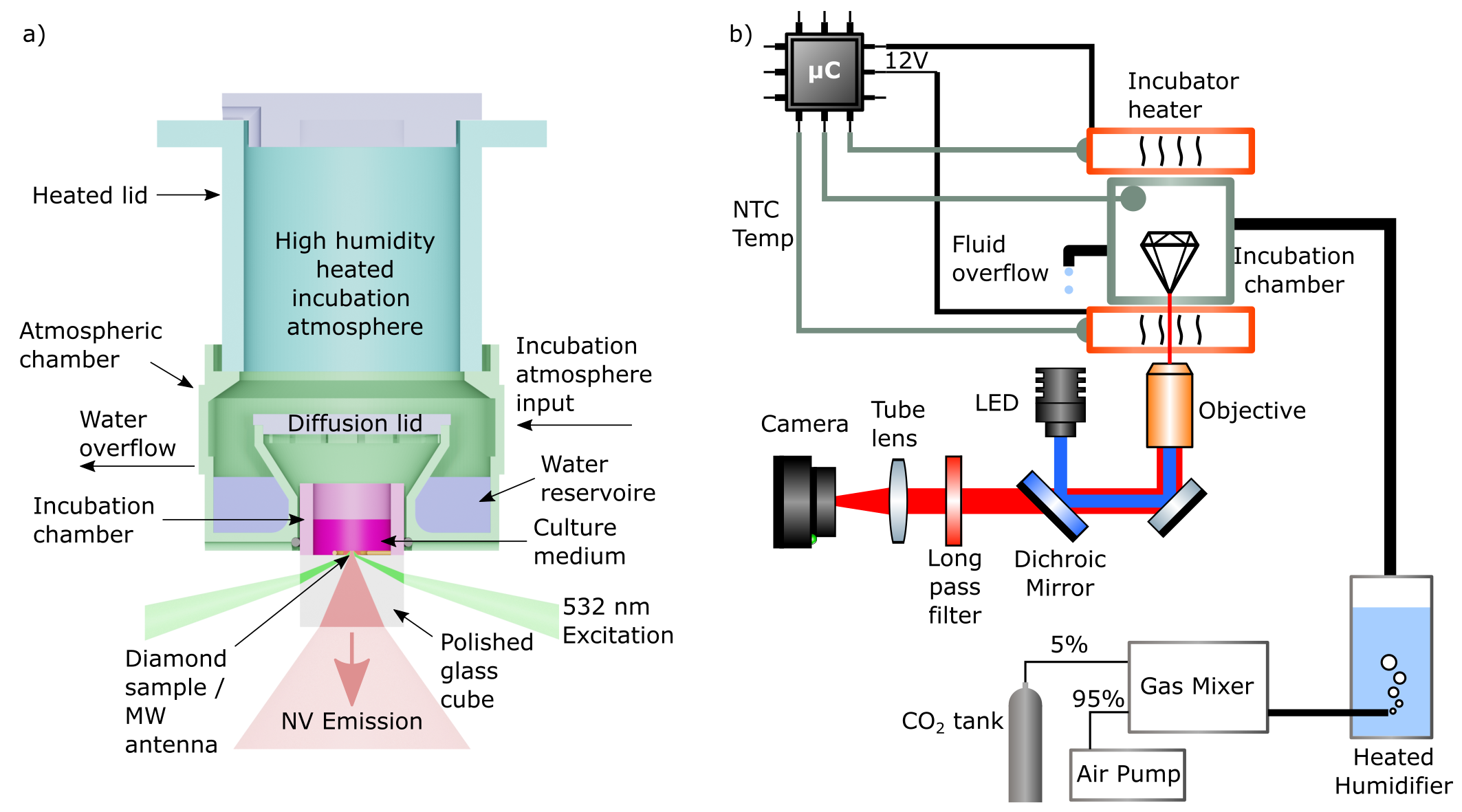}
\caption{a)
Cross-section render of the incubation chamber mounted in the NV widefield
magnetometry setup. The diamond sample (red) is mounted on a polished glass cube
(grey). The incubation volume defined by an acrylic cylinder (purple) is sealed
by the incubation chamber (green) and lid (blue). The sample holder (orange) and
lid are heated to provide temperature control of the incubation volume. The
incubation chamber provides ports for the gas atmosphere and a water reservoir
to maintain high levels of humidity, preventing medium evaporation. b) A full
schematic of the experimental hardware including all components required for NV
widefield magnetometry measurements and incubation chamber functionality. The
incubation chamber is connected to a gas mixer and humidifier to provide a
stable CO$_2$ atmosphere with high humidity. Temperature control is achieved by
a microcontroller providing PID control of the heating elements based on
feedback from temperature sensors on the heating elements and inside the
incubation atmosphere. A blue LED is integrated into the emission path via a
dichroic mirror to enable fluorescence imaging of biological samples. The
long-pass filter is exchanged for a bandpass filter to isolate fluorescence
signals during imaging.}
\label{fig:2}
\end{figure}

To provide a stable environment for long-term cultivation of cells inside an NV
experiment, we designed a quantum sensing incubator compatible with the NV
widefield magnetic field microscope (illustrated in \autoref{fig:2}a).
The incubation chamber is constructed from an acrylic cylinder glued to the
glass cube via optical adhesive (NOA61, Norland) \cite{Le_Sage2013-at}. The
chamber is carefully sealed to prevent leakage of cell culture medium,
especially around the microwave wire feed through. A rubber gasket is placed
around the incubation chamber and the atmospheric chamber is placed above the
cylinder to provide a closed off environment for biological samples.

The incubator provides temperature control by a heated aluminum holder, which
positions the incubator above the objective, as well as heating the incubator
lid \SIrange{2}{3}{\degreeCelsius} above the incubation temperature to prevent
condensation (see \autoref{fig:2}b). A microcontroller (Arduino Nano) is
employed to provide PID temperature control of the aluminum heating elements
using thick film resistors (MP825-20.0-1\%, Caddock) and negative temperature
coefficient thermistors (NTC) as temperature sensors. The process value is
controlled to be around \SI{34}{\degreeCelsius} to provide an incubation
temperature slightly below \SI{37}{\degreeCelsius} to ensure no significant
heating above this temperature in case of a beta value shift of the thermistors
due to environmental temperature changes in the laboratory. The incubation
chamber consists of a 3D printed custom-built geometry printed from a
biologically compatible material  (BioMed White Resin, Formlabs) on a
stereolithography (SLA) printer (Form 4, Formlabs). This allows full design
freedom to integrate necessary features such as ports for the gas mixture
(\SI{5}{\percent} CO$_2$ and \SI{95}{\percent} air) premixed by a gas mixer (2GF
Mixer, Okolabs) and water reservoir to maintain oversaturated humidity to
prevent evaporation and the subsequent change in pH value of the small volume of
cell culture medium (as illustrated in \autoref{fig:2}b). The air mixture is
humidified in a temperature controlled humidifier (CO2-500ml, Bioscience Tools)
and subsequently directed into the incubation chamber. Though condensation of
the humidity in the gas mixture inside the supply line is unavoidable (however
it could be diminished by heating the line to \SI{37}{\degreeCelsius}), we
included an overflow inside the incubation chamber. The resulting condensate
replenishes the internal water reservoir, providing a stable humidity level.
Internal separation of the cultivation volume and the incubation atmosphere by a
lid (above the acrylic cylinder illustrated in \autoref{fig:2}a) covering the
acrylic cylinder prevents excessive evaporation of the culture medium. Grooves
on the incubation volumes top edge allow limited gas exchange to enable CO$_2$
diffusion from the gas mixture into the incubation volume. This provides the
necessary \SI{5}{\percent} CO$_2$, while preventing unnecessary evaporation of
the culture medium. The full incubation chamber fits between the bias magnets
(see \autoref{fig:1}a) to provide a long-term stable environment without
degrading material due to the high humidity or interfering with ODMR measurement
requirements. Temperature sensors at both heaters and in the incubation
atmosphere provide real-time temperature monitoring. Temperature control is
achieved by calibrating the temperature gradient from the sensor on the outside
heater (orange sample holder in \autoref{fig:1}a) to the incubation chamber
interior at the position of the diamond. Calibration was done with deionized
water in the incubation chamber and a temperature sensor on the diamond surface.
A stable temperature gradient of \SI{3}{\degreeCelsius} between the aluminum
heater and the diamond surface was determined. This allows indirect temperature
control without a sensor present inside the chamber, as this would interfere
with the ODMR measurement. In addition to the temperature calibration the
influence of the microwave on the thermal environment inside the incubation
volume was probed. It was found that the employed microwave power did not cause
any measurable heating in both DI water and culture medium. This
geometry allows the hardware components to be placed outside the incubation
atmosphere, preventing contamination or degradation of sensitive components from
the high humidity, while still providing a sterile and stable environment for
biological samples. The incubation volume could be manufactured in total from
the biocompatible resin, but to simplify manufacturing and improve reusability,
we opted for a hybrid design instead. For a detailed description of the
incubator design, as well as technical drawings and CAD illustrations, the
authors refer the reader to the SI.

To validate the functionality of the incubation chamber, we cultured HT29 human
colorectal cancer cells on a microscope cover slide functionalized with
Fibronectin for a total of 90 hours. The cover slide was glued to the
bottom of the incubator to seal the chamber. For all incubation experiments, a
single cell suspension was drop cast onto the cover slide filling the incubation
chamber with \SI{500}{\mu\liter} of medium with approximately 40000 cells. For
the extended cultivation period, periodic images were captured over the course
of the experiment using the brightfield microscopy capability of the microscope.
Cell viability was confirmed by observing proliferation and cell adhesion over
the full duration, with proliferation events still occurring after multiple days
of incubation inside the chamber under constant conditions. The experiment was
repeated using the glass cube and diamond setup as the surface for the cells to
confirm cell viability directly on the diamond sample. The cells however showed
difficulties adhering to the bare diamond surface. Functionalizing the sample
surface with either Fibronectin or Poly-L-Lysine (PLL) improved adhesion, which
was confirmed in subsequent experiments with 48 hours of continuous incubation
and imaging. Both approaches showed good cell adhesion, while Fibronectin
provided a more natural extracellular matrix for the cells. Due to the faster
initial cell adhesion of PLL's electro-static mechanism, we opted for PLL
surface functionalization for subsequent experiments to ensure sufficient
adhesion before the start of the incubation period. A more detailed comparison
of both adhesion layers is provided in the SI.

\begin{figure}[H]
\centering
\includegraphics[width=\linewidth]{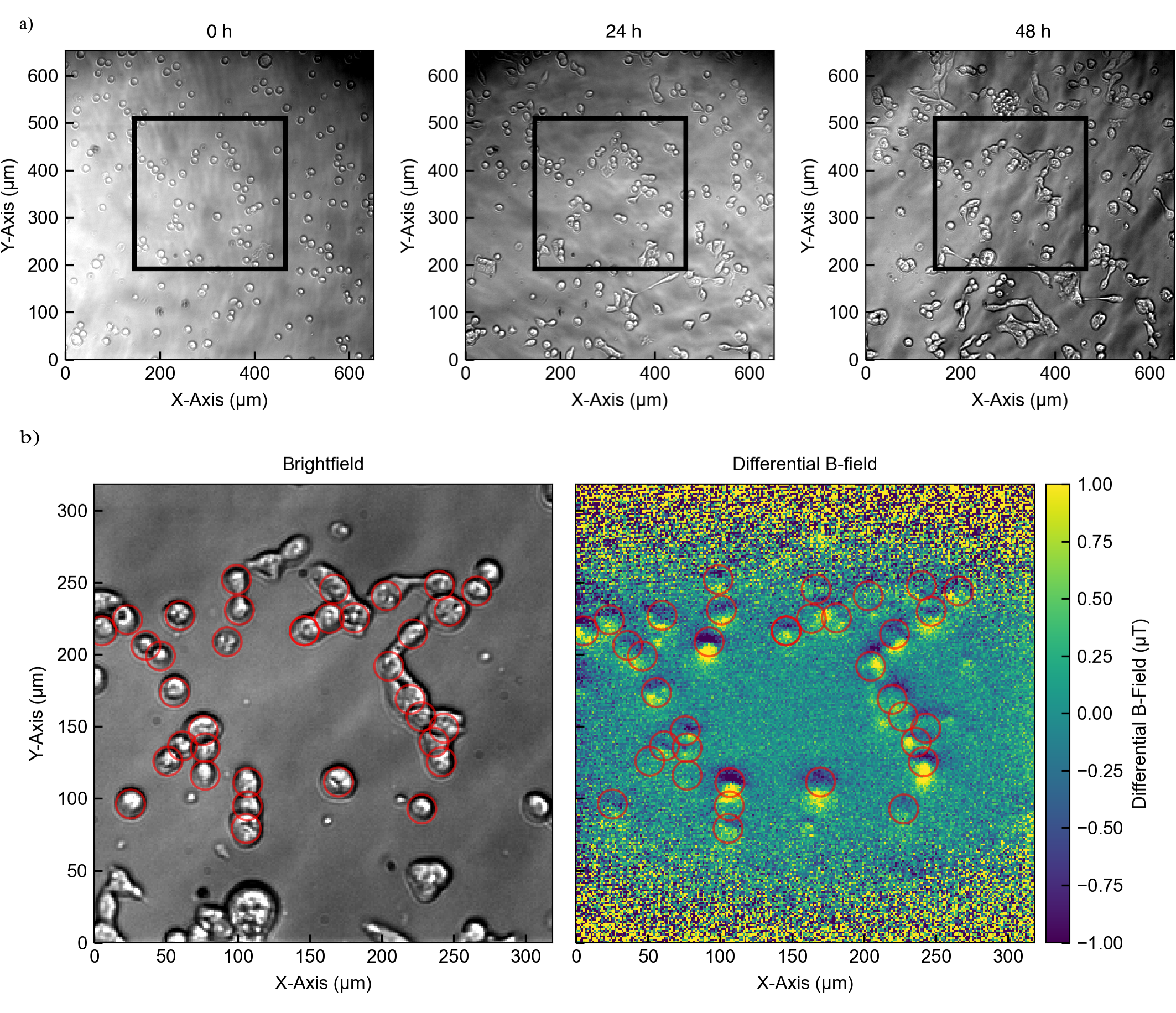}
\caption{a) Brightfield images at the start, intermediate and end times of a 48
hour incubation of magnetically labeled HT29 cells on a PLL treated diamond
surface inside the incubation chamber of the NV widefield magnetometry setup. The black frame
indicates the magnetometry FoV. The cells show typical morphology and
proliferation behavior over the course of the experiment. b) Corresponding
magnetic field image, after background subtraction, of the labeled cells
captured at the end of the incubation period. The dipole patterns generated by
the SPIONs on the cell surface are clearly visible, demonstrating successful
magnetic labeling and imaging of the cells after long-term incubation. The
magnetic field labeling homogeneity is diminished due to morphology changes and
proliferation of the cells during incubation (see SI). The upper and lower edge of the
image appears noisy, as the excitation intensity decays towards the edges of the
FoV. The gaussian beam profile is projected onto the diamond surface under the
TIR angle, resulting in an elongation along the x-axis, producing an elliptical
magnetic field FoV.}
\label{fig:3}
\end{figure}

The improved adhesion allowed for successful long-term cultivation on the
diamond surface, as shown in \autoref{fig:3}a. The HT29 cells,
magnetically labeled with SPIONs as outlined in previous work in more detail in
\cite{Pointner2025-acs}, showed typical morphology and proliferation behavior
over the course of the experiment. The cells remained viable and proliferated
over the full duration of 48 hours, confirming the suitability of the
incubation chamber for long-term cultivation directly on the diamond surface.
Subsequent magnetic field imaging of the labeled cells in \autoref{fig:3}b
showed clear dipole patterns observing expected deformations in the local
magnetic field generated by the SPIONs because of morphology changes and
proliferation (see SI for brightfield image sequences of cells undergoing
morphological changes), changing the homogeneous distribution of the magnetic
labels on the cell surface.

\section{Conclusion}
\label{sec:conclusion}

We have demonstrated a purpose-built incubation platform specifically designed
for long-term cultivation of biological samples on NV widefield magnetometry
platforms. The compact incubation chamber successfully maintains a stable
physiological environment for cells while accommodating the unique geometric
constraints imposed by TIR excitation geometry advantageous for NV widefield
microscopy. Our validation experiments with HT29 colorectal cancer cells
demonstrate sustained cell viability and proliferation over extended periods of
90 hours, confirming that the incubation system provides adequate environmental
control for both cell culture applications. The system maintains precise
temperature control, regulated CO$_2$ atmosphere, and stable humidity levels
while keeping measurement components such as the magnets, mechanical stages and
optics isolated from the incubation environment. Surface functionalization with
Fibronectin or PLL proved essential for cell adhesion to the diamond sensing
surface, with Fibronectin providing superior proliferation rates due to its more
physiologically relevant extracellular matrix properties. The hybrid design
approach balancing 3D printed biocompatible components with reusable elements
offers a practical solution that reduces cost while maintaining the flexibility
needed for the NV widefield magnetometry platform. The presented design provides
a life-sustaining environment while still enabling high-quality ODMR
measurements of biological samples.

Future improvements to this platform could include real-time CO$_2$ monitoring
to provide closed-loop feedback control of gas composition, implementation of
perfusion capabilities to enable extended cultivation periods beyond 90 hours
while maintaining optimal nutrient supply and waste removal, and scaling to
multi-well configurations for parallel cultivation of multiple samples under
identical environmental conditions. These enhancements would further expand the
applicability of this incubation system for complex biological investigations
on NV widefield magnetometry platforms requiring long-term cultivation of
cells.

\medskip
\textbf{Acknowledgements} \par Roland Nagy was supported by the
Deutsche
Forschungsgemeinschaft (DFG) NA1764/2-1 and INST 90/1252-1 FUGG, as
well as BMBF
(QUBIS). Regine Schneider-Stock was supported by the Deutsche
Forschungsgemeinschaft (DFG) SCHN477/19-1 and BMBF (QUBIS). Rainer Tietze was
supported by the Deutsche Forschungsgemeinschaft (DFG) TI1174/2-1, BMBF (QUBIS)
and Hans Wormser, Herzogenaurach, Germany. Christina Janko was funded by the
Professorship for AI-Guided Nanomaterials within the framework of the Hightech
Agenda (HTA) of the Free State of Bavaria. The authors acknowledge the use of
Claude 4.1 Opus (accessed November 2025) to support creation of the abstract
and
to assist with sentence structure adjustments, and translation of selected
phrases. All AI-generated suggestions were reviewed, revised, and approved by
the authors, who take full responsibility for the accuracy and integrity of the
work.

\bibliographystyle{unsrt}
\bibliography{references}

\begin{thebibliography}{10}

\bibitem{Mullassey2008-xx}
D~Mullassey, C~A Horton, C~D Wood, and M~R~H White.
\newblock Single live cell imaging for systems biology.
\newblock {\em Essays in Biochemistry}, 45:121--133, 2008.

\bibitem{Gahlmann2014-sr}
Andreas Gahlmann and W~E Moerner.
\newblock Exploring bacterial cell biology with single-molecule tracking and
  super-resolution imaging.
\newblock {\em Nature Reviews Microbiology}, 12(1):9--22, 2014.

\bibitem{Schmidt2021-mf}
Roman Schmidt, Tobias Weihs, Christian~A Wurm, Isabelle Jansen, Jasmin Rehman,
  Steffen~J Sahl, and Stefan~W Hell.
\newblock {MINFLUX} nanometer-scale {3D} imaging and microsecond-range tracking
  on a common fluorescence microscope.
\newblock {\em Nature Communications}, 12:1478, 2021.

\bibitem{Lu2023-dm}
Dengyun Lu, Guoshuai Zhu, Xing Li, Jianyun Xiong, Danning Wang, Yang Shi, Ting
  Pan, Baojun Li, Luke~P Lee, and Hongbao Xin.
\newblock Dynamic monitoring of oscillatory enzyme activity of individual live
  bacteria via nanoplasmonic optical antennas.
\newblock {\em Nature Photonics}, 17:904--911, 2023.

\bibitem{Zhang2021Toward}
Tongtong Zhang, G.~Pramanik, Kaiwen Zhang, Michal Gulka, Lingzhi Wang, J.~Jing,
  Feng Xu, Zifu Li, Q.~Wei, P.~C{\'i}gler, and Zhiqin Chu.
\newblock Toward quantitative bio-sensing with nitrogen-vacancy center in
  diamond.
\newblock {\em ACS Sensors}, 6(6):2077--2107, 2021.

\bibitem{Hollendonner2023-yi}
M~Hollendonner, S~Sharma, S~K Parthasarathy, D~B~R Dasari, A~Finkler, S~V
  Kusminskiy, and R~Nagy.
\newblock Quantum sensing of electric field distributions of liquid
  electrolytes with {NV-centers} in nanodiamonds.
\newblock {\em New Journal of Physics}, 25(9):093008, 2023.

\bibitem{Liu2022-xx}
Kristina~S Liu, Alex Henning, Markus~W Heindl, and Dominik~B Bucher.
\newblock Surface {NMR} using quantum sensors in diamond.
\newblock {\em Proceedings of the National Academy of Sciences},
  119(5):e2111607119, 2022.

\bibitem{Schirhagl2014-xx}
Romana Schirhagl, Kevin Chang, Michael Loretz, and Christian~L Degen.
\newblock Nitrogen-vacancy centers in diamond: nanoscale sensors for physics
  and biology.
\newblock {\em Annual Review of Physical Chemistry}, 65:83--105, 2014.

\bibitem{Kucsko2013-th}
G~Kucsko, P~C Maurer, N~Y Yao, M~Kubo, H~J Noh, P~K Lo, H~Park, and M~D Lukin.
\newblock Nanometre-scale thermometry in a living cell.
\newblock {\em Nature}, 500(7460):54--58, 2013.

\bibitem{PeronaMartinez2020-nd}
Felipe Perona~Mart{\'\i}nez, Anggrek~Citra Nusantara, Mayeul Chipaux,
  Sandeep~Kumar Padamati, and Romana Schirhagl.
\newblock Nanodiamond relaxometry-based detection of free-radical species when
  produced in chemical reactions in biologically relevant conditions.
\newblock {\em ACS Sensors}, 5(12):3862--3869, 2020.

\bibitem{Fan2025-iv}
Siyu Fan, Yue Zhang, Anna~P Ainslie, Ren{\'e}e Seinstra, Tao Zhang, Ellen
  Nollen, and Romana Schirhagl.
\newblock In vivo nanodiamond quantum sensing of free radicals in
  \textit{Caenorhabditis elegans} models.
\newblock {\em Advanced Science}, 2025.

\bibitem{Pointner2025-acs}
Andre Pointner, Daniela Thalheim, Sarah Belasi, Lukas Heinen, Cristian Bonato,
  Tobias Luehmann, Jan Meijer, Rainer Tietze, Christoph Alexiou, Regine
  Schneider-Stock, and Roland Nagy.
\newblock Optimizing {SPION} labeling for single-cell magnetic microscopy.
\newblock {\em The Journal of Physical Chemistry Letters}, 16(30), 2025.

\bibitem{Le_Sage2013-at}
D~Le~Sage, K~Arai, D~R Glenn, S~J DeVience, L~M Pham, L~Rahn-Lee, M~D Lukin,
  A~Yacoby, A~Komeili, and R~L Walsworth.
\newblock Optical magnetic imaging of living cells.
\newblock {\em Nature}, 496(7446):486--489, 2013.

\bibitem{Chen2022-cs}
Sanyou Chen, Wanhe Li, Xiaohu Zheng, Pei Yu, Pengfei Wang, Ziting Sun, Yao Xu,
  Defeng Jiao, Xiangyu Ye, Mingcheng Cai, Mengze Shen, Mengqi Wang, Qi~Zhang,
  Fei Kong, Ya~Wang, Jie He, Haiming Wei, Fazhan Shi, and Jiangfeng Du.
\newblock Immunomagnetic microscopy of tumor tissues using quantum sensors in
  diamond.
\newblock {\em Proceedings of the National Academy of Sciences},
  119(5):e2118876119, 2022.

\bibitem{Glenn2015-gy}
D~R Glenn, K~Lee, H~Park, R~Weissleder, A~Yacoby, M~D Lukin, H~Lee, R~L
  Walsworth, and C~B Connolly.
\newblock Single-cell magnetic imaging using a quantum diamond microscope.
\newblock {\em Nature Methods}, 12(8):736--738, 2015.

\bibitem{Friedrich2022-sf}
Ralf~P Friedrich, Mona Kappes, Iwona Cicha, Rainer Tietze, Christian Braun,
  Regine Schneider-Stock, Roland Nagy, Christoph Alexiou, and Christina Janko.
\newblock Optical microscopy systems for the detection of unlabeled
  nanoparticles.
\newblock {\em International Journal of Nanomedicine}, 17:2139--2163, 2022.

\bibitem{Sengottuvel2022-xx}
Saravanan Sengottuvel, Mariusz Mrózek, Mirosław Sawczak, Maciej~J Głowacki,
  Mateusz Ficek, Wojciech Gawlik, and Adam~M Wojciechowski.
\newblock Wide-field magnetometry using nitrogen-vacancy color centers with
  randomly oriented micro-diamonds.
\newblock {\em Scientific Reports}, 12:17997, 2022.

\bibitem{Wijewardhane2024-lc}
Neshika Wijewardhane, Ana~Rubio Denniss, Matthew Uppington, Helmut Hauser,
  Thomas~E Gorochowski, Eugenia Piddini, and Sabine Hauert.
\newblock Long-term imaging and spatio-temporal control of living cells using
  targeted light based on closed-loop feedback.
\newblock {\em Journal of Microbio Robotics}, 12(1):2, 2024.

\bibitem{Worcester2025-st}
Michael Worcester, Shayan Nejad, Pratyasha Mishra, Quintin Meyers, Melissa
  Gomez, and Thomas Kuhlman.
\newblock A low-cost stage-top incubation device for live human cell imaging
  using rapid prototyping methods.
\newblock {\em AIMS Biophysics}, 12(2):164--173, 2025.

\bibitem{quantum-diamonds}
{QuantumDiamonds GmbH}.
\newblock Friedenstraße 18, 81671 münchen, germany, 2025.
\newblock Accessed: January 2025.

\bibitem{Bucher2019-nmr}
Dominik~B Bucher, Diana P~L Aude~Craik, Mikael~P Backlund, Matthew~J Turner,
  Oren Ben~Dor, David~R Glenn, and Ronald~L Walsworth.
\newblock Quantum diamond spectrometer for nanoscale {NMR} and {ESR}
  spectroscopy.
\newblock {\em Nature Protocols}, 14:2707--2747, 2019.

\bibitem{Robertson2019-bl}
Julia Robertson, Cushla McGoverin, Fr{\'e}d{\'e}rique Vanholsbeeck, and Simon
  Swift.
\newblock Optimisation of the protocol for the {LIVE/DEAD} {BacLight} bacterial
  viability kit for rapid determination of bacterial load.
\newblock {\em Frontiers in Microbiology}, 10:801, 2019.

\end{thebibliography}

\end{document}